\documentclass{iopart}
\usepackage{epsfig}
\usepackage{times}
\newcommand{\Label}[1]{\label{#1}}
\newcommand{\BEC}{Bose-Einstein condensate}

\newcommand{\ccfs}{cold collision frequency shift}

\begin{document}

\bibliographystyle{iopacm}

\title[Analysis of $g_2$ in hydrogen\dots]
{Analysis of {$g_2$} for the \ccfs\
in the hydrogen condensate experiments}
\author{C.W. Gardiner and A.S. Bradley}
\address{School of Chemical
and Physical Sciences, Victoria University, Wellington, New
Zealand}
\begin{abstract}
We compute an approximate set of longitudinal quasiparticle modes
for a hydrogen condensate as produced in the MIT experiments. An
expansion in quasiparticles using a simple one-dimensional
Bogoliubov picture shows however that at the high temperatures ($
\approx 44\mu{\rm K}$) and in the very shallow trap employed ($
\omega_z = 2\pi\times 10.2 {\rm Hz}$) the contribution to the
density from the quasiparticles is about 20\% of that from the
condensate mode, leading to an effective $ g_2({\bf x},{\bf x})$
which varies between 1 and 3 depending on the position in the
condensate.
\end{abstract}
\section{Introduction}
The analysis of the experimental data on a spin-polarized atomic
hydrogen \BEC\, as given in
\cite{Killian1998ax,Fried1998a,KillianThesis,FriedThesis}, relates
the density of the condensate or of the non-condensed gas to the
frequency for the two-photon 1s--2s transition by the cold
collision line-shift formula
\begin{eqnarray}\Label{z1}
\Delta\nu_{\rm 1s-2s} = (a_{\rm 1s-2s}-a_{\rm 1s-1s})
{2\hbar g_2({\bf x},{\bf x}) n_{\rm 1s}\over m}.
\end{eqnarray}
In this formula $ a_{\rm 1s-1s} $ and $ a_{\rm 1s-2s}$ are respectively the
scattering lengths for collisions between the two atoms in the 1s state or of
one atom in each state, $ n_{\rm 1s}$ is the density of the 1s atoms, and
\begin{eqnarray}\Label{z2}
g_2({\bf x},{\bf x}') &=&
{\langle\psi^\dagger({\bf x})\psi^\dagger({\bf x'})
\psi({\bf x'})\psi({\bf x})\rangle\over
 n_{\rm 1s}({\bf x})n_{\rm 1s}({\bf x}')}.
\end{eqnarray}
For a non-condensed thermal gas $ g_2({\bf x},{\bf x})=2$, while for a
pure condensate $ g_2({\bf x},{\bf x})=1$.

The derivation of the formula (\ref{z1}) is given in
\cite{KillianThesis}, though it is acknowledged there that this derivation 
is only valid for special cases; that is for a
spatially homogeneous thermal non-condensed gas, and for a pure
condensate. Intermediate situations, such as those which appear to
pertain in practice have not been treated, and indeed are very difficult to analyze.

In the experiments with the choice of $g_2({\bf x},{\bf x})=1$, frequency
shifts measured in the \BEC\ experiments would correspond to densities
too high to be credible, and the data were analyzed using $ g_2({\bf
x},{\bf x})=2$.  This was justified by the manifest high temperature
of the system, and the conjecture that condensation occurred into a
multiplicity of states.

This paper is the first of two on this subject. Here we shall concern 
ourselves specifically with the computation of $g_2$  for the case of hydrogen 
condensate with a significant thermal component, in order to demonstrate 
what would be predicted by its use if the formula (\ref{z1}) is 
assumed to be valid in that case.
In our second paper \cite{newpaper} we shall show that the application 
of (\ref{z1}) to a condensate with a thermal component is not 
justified, and we shall give  a full dynamic formulation of the actual 
processes involved.

The evaluation of $g_2$ is relevant to other phenomena, such as 
two-body loss processes, so its computation is still of value, even if 
it is not directly applicable to the analysis of the \ccfs\ in these 
experiments. Furthermore, in order to do the computations
of our second paper, we shall need the eigenfunctions computed here.
 We will use a simple
Bogoliubov model, though it will become apparent that there should
still be significant corrections to this at the temperatures and 
densities found in practice.  A more detailed study is left for a 
later paper.

One might expect that in
general $g_2({\bf x},{\bf x})$ lies between 1 and 2, but in fact the
only {\em fundamental} limit is that $0\le g_2({\bf x},{\bf
x})\le\infty$.
However, in the case of Gaussian statistics, we get a stronger limit.
Thus in the case of
a fully thermalized gas, we can split the correlation function up using
Gaussian factorization
\begin{eqnarray}\label{1}
\langle\psi^\dagger({\bf x})\psi^\dagger({\bf x}')\psi({\bf
x}')\psi({\bf x})\rangle &=&\langle\psi^\dagger({\bf x})\psi({\bf
x})\rangle\langle\psi^\dagger({\bf x}')\psi({\bf x}')\rangle
\nonumber\\
&&+\langle\psi^\dagger({\bf x})\psi({\bf
x}')\rangle\langle\psi^\dagger({\bf x}')\psi({\bf x})\rangle
\nonumber\\
&&+\langle\psi^\dagger({\bf x})\psi^\dagger({\bf
x}')\rangle\langle\psi({\bf x}')\psi({\bf x})\rangle
\nonumber\\
\end{eqnarray}
The third term in the case of a thermal gas is zero since there is no anomalous
average
\begin{eqnarray}\label{2}
\langle\psi^\dagger({\bf x})\psi^\dagger({\bf x}')\rangle =0,
\end{eqnarray}
and setting $ {\bf x}={\bf x}'$ gives
\begin{eqnarray}\label{3}
g_2({\bf x},{\bf x}) = {\langle\psi^\dagger({\bf x})\psi^\dagger({\bf
x})\psi({\bf x})\psi({\bf x})\rangle\over \langle\psi^\dagger({\bf
x})\psi({\bf x})\rangle\langle\psi^\dagger({\bf x})\psi({\bf
x})\rangle}\to 2 .
\end{eqnarray}
Thus $ g_2({\bf x},{\bf x})=2$ is a consequence of the Gaussian nature
of the fluctuations and of the absence of an anomalous average.  The
anomalous average {\em does} have an upper bound of approximately
\begin{eqnarray}\label{4}
|\langle\psi({\bf x})\psi({\bf x}')\rangle |
 < \langle\psi^\dagger({\bf x})\psi({\bf x})\rangle
\end{eqnarray}
which gives the bound
\begin{eqnarray}\label{5}
g_2({\bf x},{\bf x}) < 3 .
\end{eqnarray}
We will find in our calculation that in fact $g_2({\bf x},{\bf x})>2$
can occur in some regions of the Hydrogen vapor-condensate system for
the experimentally realized parameters.

\section{Second-quantized Hamiltonian}
For the case under consideration the
second-quantized Hamiltonian can be written in the form
\begin{eqnarray}\Label{a1}
H &=& \int d^3{\bf x}\, \left\{
\psi^\dagger\left(T + V({\bf x}) \right)\psi
+ {u\over 2}\psi^\dagger\psi^\dagger\psi\psi\right\}
.
\end{eqnarray}
\begin{itemize}
\item[i)]
The kinetic energy operator is $ T = -\hbar^2\nabla^2/2m$
\item[ii)]
 The interaction coefficient is given in terms of the scattering length by
\begin{eqnarray}\Label{a101}
u &=& {4\pi a_{\rm 1s-1s}\hbar^2/m}
\end{eqnarray}
Experimentally
\begin{eqnarray}\Label{a102}
a_{\rm 1s-1s} &=& 0.0648{\rm nm}
\\
a_{\rm 1s-2s} &=& -1.4\pm 0.3{\rm nm}.
\end{eqnarray}
\end{itemize}

\section{Evaluation of \mbox{\lowercase{$ g_2({\bf x},{\bf x})$}} for a warm condensate}
The hydrogen condensate at MIT is formed at $ 44\mu{\rm K}$, and the radial and
axial trap frequencies are
\begin{eqnarray}\Label{c1a}
\omega_r &=& 2\pi\times 3.9{\rm kHz}
\\
\omega_z &=& 2\pi\times 10.2{\rm Hz}.
\end{eqnarray}
The ratio of these $ \omega_r/\omega_z=382$ gives the trap aspect ratio
$ l_z/l_r$, which at nearly 400:1 gives  an almost one-dimensional condensate.

As well as this, the ratios of the energies of the transverse and longitudinal
quanta to the temperature are
\begin{eqnarray}\Label{c1b}
 kT/\hbar\omega_r & \approx& 1/250
\\
kT/\hbar\omega_z & \approx& 1/10^5.
\end{eqnarray}
We can therefore expect that the average number of quasiparticles in the
longitudinal mode is of the order of $ 10^5$ for the lower levels. This does
not mean that the condensate is very different in size and shape, but its
coherence may be affected by these large excitations.  For example, the
longitudinal Kohn mode corresponds to center of mass oscillations of the whole
condensate, of mass $ 10^9 m$, and using statistical mechanics, this means that
\begin{eqnarray}\Label{c2}
\sqrt{\langle z^2\rangle} \approx 3\times10^{-4}{\rm mm}
 \approx 6\times 10^{-6} l_z,
\end{eqnarray}
which is a completely unobservable deviation.

\subsection{Approximate evaluation of mode functions}
In order to get an approximate idea of the thermal fluctuations we will
compute
the wavefunctions of the lower lying excitations using a hydrodynamic
approximation based on the work of Zaremba \cite{Zaremba1998b}.  This method
gives an approximate form for the full three dimensional wavefunction in the
case that the radial modes are not excited; thus it is possible do a
calculation by means of a one dimensional wave equation, and using the
Thomas-Fermi approximation for the condensate wavefunction this give
eigenfunctions analytically in terms of Jacobi polynomials.

\subsubsection{Hydrodynamic equations}
 The system
can be characterized by a density $ \rho({\bf x},t)$ and a phase $
\phi({\bf x},t)$, and the hydrodynamic equations are written in
terms of the linearized density fluctuation $ \delta\rho({\bf
x},t) = \rho({\bf x},t) -\rho_0({\bf x})$ as
\begin{eqnarray}\Label{c4}
\delta\dot\rho({\bf x},t) &=&
-{\hbar\over m}\nabla\cdot\left[{\rho_0({\bf x})}{\nabla\phi({\bf x},t)}
\right]
\\ \Label{c5}
\dot\phi({\bf x},t) &=& -{u\over\hbar}\delta\rho({\bf x},t).
\end{eqnarray}
Zaremba's analysis considers only the situation in which there is a
cylindrically symmetric condensate with no $ z$-dependence, and he shows that
in this case the speed of sound along the $ z$ direction is half that expected
in a homogeneous condensate with density equal to the peak density of the
cylindrical condensate.  We will generalize his result to the situation in
which the trap is harmonic in all directions, but is very weak along the $
z$-direction.  Following Zaremba's lead, we look for the equation of motion
for
a perturbation with the property
\begin{eqnarray}\Label{c501}
\delta\rho({\bf x},t) &=& \delta n(z) \mbox{ when } \rho_0({\bf x}) \ne 0,
\\
&=& 0  \qquad \mbox{ when }\rho_0({\bf x}) = 0.
\end{eqnarray}
Here it is assumed that  the Thomas-Fermi form
\begin{eqnarray}\Label{c502}
\rho_0({\bf x}) = {2\mu - m\omega_r^2r^2 - m\omega_z^2 z^2\over 2 u},
\end{eqnarray}
is used, so that (\ref{c501}) means that $ \delta n(z) $ is only non-zero
where there is a non-vanishing condensate density, i.e., when
\begin{eqnarray}\Label{c503}
 r < R(z) \equiv \sqrt{2\mu - m\omega_z^2z^2\over m\omega_r^2}.
\end{eqnarray}
When we substitute into the wave equation for $ \delta\rho$ that arises from
combining
(\ref{c4}) and(\ref{c5}), and then integrate over $ 2\pi r\,dr$, the resulting
equation can be written in terms of the scaled length
\begin{eqnarray}\Label{c6}
z = h \sqrt{2\mu\over m\omega_z^2}\equiv \lambda h
\end{eqnarray}
as
\begin{eqnarray}\Label{c601}
{\partial^{2}\delta n \over\partial   t^{2}} &=& {\omega_z^2\over
4}\left\{(1-h ^2){\partial^{2}\delta n \over\partial   h ^{2}} -
4h {\partial \delta n \over\partial h }\right\}.
\end{eqnarray}
The operator on the right hand side is that of the Jacobi
polynomials $ P^{(1,1)}_n(h )$, leading to the eigenvalue spectrum
\begin{eqnarray}\Label{c602}
\omega_n &=& {\omega_z\over 2}\sqrt{n(n+3}).
\end{eqnarray}
The more exact three dimensional analysis of Fliesser {\em et al.}
\cite{Fliesser1997b} gives, in
the case of an extended cigar-shaped condensate, the same formula.

Note that these Jacobi polynomials are orthogonal in the sense that
\begin{eqnarray}\Label{c603}
\int_{-1}^{1}dh (1-h ^2)P^{(1,1)}_n(h )P^{(1,1)}_m(h ) &=&
{8(n+1)\delta_{n,m}\over (2n+3)(n+2)},
\nonumber\\
\end{eqnarray}
so that the appropriate weight function is $ (1-h ^2)$, which is
proportional to the stationary condensate density.
\subsubsection{Three dimensional interpretation of the wavefunctions}
As in Zaremba's analysis, these eigenfunctions are {\em constant} in the radial
direction to the edge of
the Thomas-Fermi condensate (\ref{c502}), where they abruptly drop to zero.
This means that the excitations are all essentially sound waves travelling along
the length of the condensate.

\subsubsection{Quasiparticle wavefunctions}
We can thus write
\begin{eqnarray}\Label{c9}
\delta n(z,t) &=& \sum_{n}A_n\cos(\omega_nt)P^{(1,1)}_n(h )
\\ \Label{c10}
\phi(z,t)  &=& -\sum_n
{uA_n\over\hbar\omega_n}\sin(\omega_nt)P^{(1,1)}_n(h ).
\end{eqnarray}
The resulting condensate wavefunction $ \xi$ (normalized to the total number of
particles, not to 1) is given by
\begin{eqnarray}\Label{c11}
&&\xi({\bf x},t) =
 \sqrt{\rho_0({\bf x}) + \delta\rho({\bf x},t)}\exp(i\phi(z,t))
\\ \Label{c12}
&&\qquad\approx
\sqrt{\rho_0({\bf x})} + {\delta\rho({\bf x},t)\over 2\sqrt{\rho_0({\bf x})}}
 + i\sqrt{\rho_0({\bf x})}\,\phi(z,t)
\\ \Label{c13}
&&\qquad  =\sqrt{\rho_0({\bf x})}
\nonumber \\
&&\qquad +\sum_n A_nP^{(1,1)}_n(h ) \Bigg\{{e^{i\omega_nt}\over
2\sqrt{\rho_0({\bf x})}} \left({1\over 2} -{u\rho_0({\bf
x})\over\hbar\omega_n}\right) \nonumber \\ &&
\qquad\phantom{\sum_n A_nP^{(1,1)}_n(h )} +{e^{-i\omega_nt}\over
2\sqrt{\rho_0({\bf x})}} \left({1\over 2} +{u\rho_0({\bf
x})\over\hbar\omega_n}\right)\Bigg\}
\end{eqnarray}
This means that the equivalent $ u_n$ and $ v_n$ functions which would turn up
in a quasiparticle expansion  of the field operator,
\begin{eqnarray}\Label{c1301}
\fl\psi({\bf x},t) &=&{a_0\over
\sqrt{n_c}}\Bigg\{\sqrt{\rho_0({\bf x})}+\sum_n\left\{ \alpha_n
e^{-i\omega_n t}u_n({\bf x}) +\alpha^\dagger_n e^{i\omega_n
t}v_n({\bf x})\right\}\bigg\}
\\ \Label{1302}
\fl&\equiv&{a_0\over \sqrt{n_c}}\left\{\sqrt{\rho_0({\bf
x})}+\tilde\psi({\bf x})\right\}
\end{eqnarray}
are given by
\begin{eqnarray}\Label{c14}
u_n({\bf x}) &=& {A_n\over 2\sqrt{\rho_0({\bf x})}} \left({1\over
2} +{u\rho_0({\bf x})\over\hbar\omega_n}\right)P^{(1,1)}_n(h )
\\ \Label{c15}
v_n({\bf x}) &=& {A_n\over 2\sqrt{\rho_0({\bf x})}} \left({1\over
2} -{u\rho_0({\bf x})\over\hbar\omega_n}\right)P^{(1,1)}_n(h ),
\end{eqnarray}
where $A_n$ are normalization constants to be determined.

\subsubsection{Amplitudes orthogonal to the condensate}
The correct eigenfunctions whose quantized amplitudes represent quasiparticles
are orthogonal to the condensate wavefunction.  To check this, we evaluate
\begin{eqnarray}\Label{c16}
&&\int  d^3{\bf x}\,\sqrt{\rho_0({\bf x})}u_n({\bf x}) =
\lambda\int_{-1}^{1}dh \int_{0}^{R(z)}2\pi
r\,dr\,\sqrt{\rho_0({\bf x})}
\nonumber \\
&&\qquad\times {A_n\over 2\sqrt{\rho_0({\bf x})}} \left({1\over 2}
+{u\rho_0({\bf x})\over\hbar\omega_n}\right)P^{(1,1)}_n(h )
\\ \Label{c17}
&&\qquad= {A_n\lambda\over 2}\int_{-1}^{1}dh \int_0^{R(z)}2\pi
r\,dr \left({1\over 2}+{u\rho_0({\bf
x})\over\hbar\omega_n}\right)P^{(1,1)}_n(h ).
\nonumber \\
\end{eqnarray}
We now use the Thomas-Fermi form (\ref{c502}) for the stationary condensate
density so that
\begin{eqnarray}\Label{c170001}
\int_0^{R(z)}2\pi r\,dr = \pi R(z)^2& =& {2\pi\mu(1-h ^2)\over
m\omega_r^2},
\\
\Label{c170002} \int_0^{R(z)}2\pi r\,dr\,u\rho_0({\bf x}) &=&
{\pi\mu^2\over m\omega_r^2}(1-h ^2)^2.
\end{eqnarray}
so that, using the orthogonality property (\ref{c603}), equation (\ref{c17})
becomes
\begin{eqnarray}\Label{c170003}
&&{A_n\lambda\pi\mu\over 2m \omega_r^2}\int_{-1}^{1}dh \,(1-h
^2)P^{(1,1)}_n(h ) \left\{1+ {\mu\over \hbar\omega_n}(1-h
^2)\right\} \nonumber \\ &&\qquad =
\delta_{n,2}{\lambda\pi\mu^{2}A_{n}\over 2m
\omega_r^2\hbar\omega_n} \int_{-1}^{1}dh \,(1-h ^2)^2P^{(1,1)}_2(h
).
\end{eqnarray}
The nonorthogonal part---which turns up only in the quadrupole
mode---represents an unobservable time dependent phase of the
condensate, and is correctly treated by simply subtracting the
component parallel to the condensate
\cite{Morgan1998a,MorganThesis}, equivalent to making the
replacement in the {\em second} part of the large bracket in
(\ref{c14}) and (\ref{c15}) $ P^{(1,1)}_n(h ) \to \bar
P^{(1,1)}_n(h )$, where
\begin{eqnarray}\Label{c18}
\bar P^{(1,1)}_n(h ) &=& P^{(1,1)}_n(h ) \qquad n\ne 2,
\\ \Label{c19}
\bar P^{(1,1)}_2(h )& =& {\sqrt{15\over 56}}\left(7h ^2 -
1\right),
\end{eqnarray}
where the norm of $ \bar P^{(1,1)}_2$ has been fixed to be the same as that of
$ P^{(1,1)}_2$, namely $ 6/7$.

It is also straightforward to check that all the $ \bar
P^{(1,1)}_n$ for $ n \ge 1$ are orthogonal to each other, so that
\begin{eqnarray}\fl\nonumber\Label{c20}
&& \int d^3{\bf x}[u_n({\bf x})u_m({\bf x}) - v_n({\bf x})v_m({\bf
x})]
\\ \nonumber
 &&=\lambda\int_{-1}^{1}dh {2\pi\mu(1-h ^2)\over
m\omega_r^2}{uA_n^2\over2\hbar \omega_n}\bar P^{(1,1)}_m(h
)P^{(1,1)}_n(h ) \nonumber
\\
&&= \lambda{\pi\mu\over m\omega_r^2}
{uA_n^2\over\hbar\omega_n}\delta_{n,m}{\cal R}_n
\end{eqnarray}
where
\begin{eqnarray}\Label{c2001}
{\cal R}_n &=& \left\{{8(n+1)\over(2n+3)(n+2)}
+\delta_{n,2}\left[{\sqrt{24\over35}}-{6\over 7}\right]\right\}.
\end{eqnarray}
Since the normalization is  given by
\begin{eqnarray}\Label{c21}
\int d^3{\bf x}[u_n({\bf x})u_n({\bf x}) - v_n({\bf x})v_n({\bf x})] &=&1
\end{eqnarray}
we must choose
\begin{eqnarray}\Label{c2201}
 A_n = \sqrt{\hbar\omega_nm\omega_r^2\over \lambda{\cal R}_n\pi \mu u}.
\end{eqnarray}

\subsubsection{The number of particles per quasiparticle}
Neglecting the zero point contribution, negligible in comparison to the
thermal
contribution in this case, the number of particles per quasiparticle for the
mode $ n$ is given by
\begin{eqnarray}\Label{c2202}
Q_n &=& \int d^3{\bf x}\,[u_n({\bf x})^2 +v_n({\bf x})^2]
\\
&=&A_n^2\int d^3{\bf x}\,\left\{{P^{(1,1)}_n(h )^2\over
8\rho_0({\bf x })}+ {u^2\rho_0({\bf x})\bar P^{(1,1)}_n(h
)^2\over2\hbar^2\omega_n^2} \right\}.
\nonumber\\
\end{eqnarray}
The first term involves a divergent integral, arising from the
form (\ref{c502}) for $ \rho_0({\bf x})$. This problem is a
reflection of the failure of the linearized expansion (\ref{c12})
This can be fixed by noting that the true density does not go to
zero at $ h =1$, but reaches a value dependent on the healing
length.  The result is only logarithmically divergent, and any
estimate of it is very much smaller than the second term, which
does not diverge.  We will therefore neglect this first term.

Using the expression (\ref{c502}) for $ \rho_0({\bf x})$, we need
the results
\begin{eqnarray}\Label{c2203}
\fl&&\int_{-1}^{1} dh  (1-h^2)^2[\bar P^{(1,1)}_n(h )]^2 \equiv
{\cal S}_n
\\ \Label{c2204}
\fl&&={8(n+1)\over
(2n+3)(n+2)}-{8(n+1)\over(2n+3)^2}\left\{{(n+1)(n+3)\over(n+2)(2n+5)}
+ {n\over(2n+1)}\right\}, \qquad(n\ne 2)
\\ \Label{c2205}
\fl&&= {8\over 21}, \hbox to 9.7cm{\hfil}(n= 2)
\end{eqnarray}
\begin{figure}
\hskip 26mm
\epsfig{file=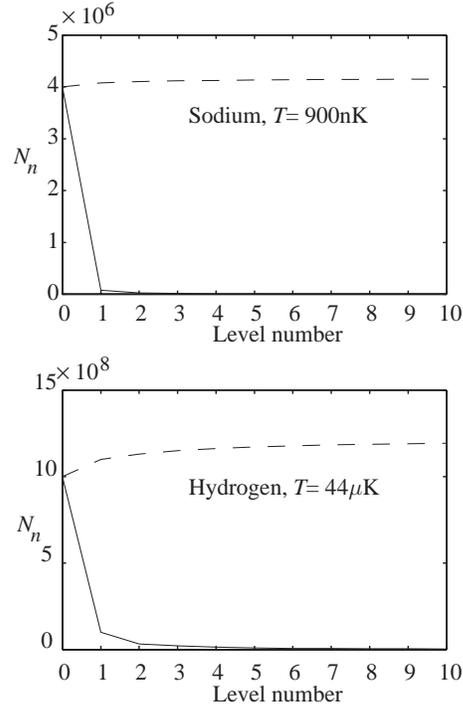,width=6cm}
\caption{Comparison of atom numbers in the lowest quasiparticle
modes for a sodium and  a hydrogen condensate. a) Solid line:
occupation of each mode; b) Dotted line: cumulative number up to
the given level. \label{fig.Q_num.eps}}
\end{figure}
and using the explicit form for $ A_n$
(\ref{c2201}), we find
\begin{eqnarray}\Label{c23}
Q_n &=& { u\bar\rho{\cal S}_n\over 2\hbar\omega_n{\cal R}_n}.
\end{eqnarray}
Here $ \bar\rho= \mu/u$ is  the peak condensate density, which in the
experiment of Fried {\em et al.} \cite{Fried1998a} was reported to be
$ 4.8\times 10^{21}{\rm m}^{-3}$.
\subsection{Numbers and densities for the experimental situation}
Using the result (\ref{c602}) and the experimental value for $
\omega_z$, we have $ \omega_n = 2\pi\times 10.2{\rm
Hz}\times\sqrt{n(n+3)/4}$, leading to $ Q_n \approx 3776{\cal
S}_n/\sqrt{n(n+3)}{\cal R}_n$.  This would lead to
\begin{eqnarray}\Label{c24}
N_{n}=Q_n{kT\over \hbar\omega_n} \approx{6.9\times10^8{\cal
S}_n\over n(n+3){\cal R}_n}
\end{eqnarray}
particles in each mode.
In Fig.\ref{fig.Q_num.eps} we illustrate the
contributions to the particle numbers from the condensate and the
first two condensate modes for the hydrogen condensate and for a
typical MIT sodium condensate \cite{Miesner1998b}, whose geometry
is a much less extreme cigar shape.  The net contribution from the
quasiparticle modes is about 2\% of the number in the condensate
mode  for sodium, but for Hydrogen approaches 20\% of the number
in the condensate mode.

Notice that convergence appears after about 10 modes.

\begin{figure}
\vbox{\parindent=2.6cm%
\epsfig{file=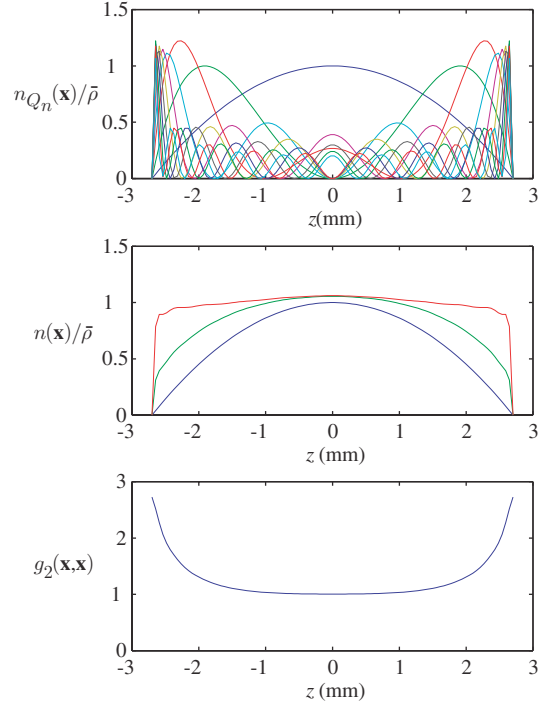,width=7cm} \caption{\Label{fig2.eps}Plots
showing: a) The individual contributions to the total density from
the condensate and the first 10 eigenfunctions; b) From bottom to
top, the condensate density, the total density, and the total
density multiplied by $ g_2({\bf x},{\bf x})$ (Normalization for
both chosen so that the peak density arising from the condensate
wavefunction is 1); c) The $ g_2({\bf x},{\bf x})$ arising from
all of the components.}}
\end{figure}

\subsubsection{Particle density arising from quasiparticles}
The contribution to $ n(h )$ arising from the quasiparticles is
simply of the form of the integrand in (\ref{c2202}) for each
mode, with the resulting overall density taking the form
\begin{eqnarray}\Label{c25}
 n_Q({\bf x}) &=& \sum_n{A_n^2kT\over \hbar\omega_n}
\left\{{P^{(1,1)}_n(h )^2\over 8\rho_0({\bf x})}+ {u^2\rho_0({\bf
x})\bar P^{(1,1)}_n(h )^2\over2\hbar^2\omega_n^2} \right\}
\end{eqnarray}
We will need the anomalous average as well, that is the term
\begin{eqnarray}\Label{c26}
m_Q({\bf x}) &=& \left\langle\tilde\psi({\bf x})^2\right \rangle
\end{eqnarray}
and this is quite readily shown to be
\begin{eqnarray}\Label{c27}
 m_Q({\bf x})  & =& \sum_n{A_n^2kT\over \hbar\omega_n}
\left\{{P^{(1,1)}_n(h )^2\over 8\rho_0({\bf x})}- {u^2\rho_0({\bf
x})\bar P^{(1,1)}_n(h )^2\over2\hbar^2\omega_n^2} \right\}
\end{eqnarray}
In the degree of approximation we are using,   in which the first
terms in the bracketed parts of the right hand sides of
(\ref{c25}) and (\ref{c27}) are neglected, it follows that $
m_Q({\bf x}) = - n_Q({\bf x})$---the anomalous term is thus very
important.

\subsubsection{Computation of $ g_2({\bf x},{\bf x})$}
The quasiparticle contribution to the field operator
\begin{eqnarray}\Label{c28}
\tilde\psi({\bf x}) = \sum_n\left\{
\alpha_n e^{-i\omega_n t}u_n({\bf x})
+\alpha^\dagger_n e^{i\omega_n t}v_n({\bf x})\right\}
\end{eqnarray}
is composed of operators $ \alpha_n,\alpha_n^\dagger$, for which the density
operator is quantum Gaussian with zero mean, so the Gaussian factorization of
the the four point correlation function is
\begin{eqnarray}\Label{c29}
\left\langle\tilde\psi^\dagger({\bf x})\tilde\psi^\dagger({\bf x})
\tilde\psi({\bf x})\tilde\psi({\bf x})\right\rangle&=&
2 n_Q({\bf x})^2 + |m_Q({\bf x})^2|
\\ \Label{c30}
&\approx& 3 n_Q({\bf x})^2.
\end{eqnarray}

Thus, for the full field operator
\begin{eqnarray}\Label{c31}
&&\left\langle\psi^\dagger({\bf x})\psi^\dagger({\bf x})
\psi({\bf x})\psi({\bf x})\right\rangle =
\rho_0({\bf x})^2
 + 4\rho_0({\bf x})
n_Q({\bf x})
\nonumber \\
&&\qquad
+ 2\rho_0({\bf x})
m_Q({\bf x})
+ 2 n_Q({\bf x})^2 + |m_Q({\bf x})^2|
\\ && \quad\approx
\rho_0({\bf x})^2
+ 2\rho_0({\bf x})
n_Q({\bf x})
+ 3 n_Q({\bf x})^2
\end{eqnarray}

Noting now that the particle density is
\begin{eqnarray}\Label{c32}
n({\bf x}) &= &{\rho_0({\bf x})} + n_Q({\bf x})
\\
&\equiv&  {\rho_0({\bf x})}[1+ q({\bf x})]
\end{eqnarray}
which defines $ q({\bf x})$, we see that
\begin{eqnarray}\Label{c33}
 g_2({\bf x},{\bf x}) &=&1 +2\left({q({\bf x})\over 1 +q({\bf x})}\right)^2.
\end{eqnarray}
Notice that the possible values of $  g_2({\bf x},{\bf x})$  are between 1 and 3, and
that
$  g_2({\bf x},{\bf x})$ does depend on position.

Using these expressions we can now compute both $ n({\bf x})$ and
$  g_2({\bf x},{\bf x})$, and
these are illustrated in Fig.\ref{fig2.eps}.

\begin{figure}
\parindent=2.6cm%
\epsfig{file=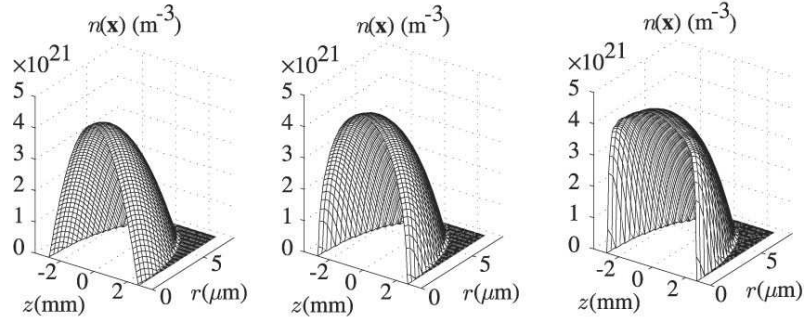,width=10.5cm}
\caption{\label{fig3.eps} %
Plots vs.  $r$ and $z$ showing: a) The condensate density (in the
Thomas-Fermi approximation);
b) The density arising from the condensate plus first 10
quasiparticle states (thermally occupied);
c) The effective density, obtained from b) by multiplying by $
g_2({\bf x},{\bf x})$}
\end{figure}%

\subsection{Interpretation}
Fig.\ref{fig3.eps} consists of three plots, to illustrate the two
major effects arising from this calculation. Comparing a) with b),
it can be seen how the condensate density is modified both in
shape and size by the excitation of the quasiparticles. Comparison
of b) with c) shows how inclusion of the factor of $ g_2({\bf
x},{\bf x})$ would modify the signal from the experiment not by
simply  multiplying by a factor of 2 as has been assumed in the
experiment.

\section{Conclusion}
The results of this calculation are only of indicative interest for the
following reasons:
\begin{itemize}
\item[a)]  The occupations of the quasiparticle levels are large enough that the
simple Bogoliubov formulation may need modification.
\item[b)] There is as yet no justification for simply including the factor of $
 g_2({\bf x},{\bf x})$ as the appropriate correction for correlations.
\item[c)]   The $g_2({\bf x},{\bf x})$ computed is the result of an
effective interpolation (given by (\ref{c33})) of $g_2({\bf
x},{\bf x})\sim 3$ for the quasiparticles and $g_2({\bf x},{\bf
x})\sim 1$ for the condensate. It can be shown in fact that the
experimental frequency shift which results in the hydrogen
condensate experiments is different for condensate and
quasiparticles, so that there is no simple interpretation in terms
of a single frequency shift determined by $g_2({\bf x},{\bf x})$.
\end{itemize}
These matters are be attended to in our second paper
\cite{newpaper}, but it is already clear that the correct
interpretation of the cold-collision shift in these experiments is
a matter of great significance.

\ack
We wish to thank Dan Kleppner, Tom Greytak, Stephen Moss, Lorenz Willman and 
Kendra Vant for useful discussions about the hydrogen condensate experiments 
and hospitality in {MIT}. This research was
supported by the Royal Society of New Zealand under the Marsden
Fund Contract PVT-902.
\section*{References}
\bibliography{Hyd1}

\end{document}